\documentclass{emulateapj}
\usepackage{natbib,epsfig}
\newcommand{\beq}{\begin{equation}}
\newcommand{\eeq}{\end{equation}}
\newcommand{\bea}{\begin{eqnarray}}
\newcommand{\eea}{\end{eqnarray}}

\newcommand{\rem}[1]{ }
\bibpunct[,]{(}{)}{;}{a}{}{,}

\shorttitle{Nonlinear saturation of the streaming instability}
\shortauthors{Gargat\'e et al.}

\begin{document}

\title{The nonlinear saturation of the non-resonant kinetically driven  streaming instability}

\author{L. Gargat\'e\altaffilmark{1}}
\affil{GoLP/Instituto de Plasmas e Fus\~ao Nuclear, Instituto Superior T\'ecnico, Av. Rovisco Pais, 1049-001 Lisbon, Portugal}
\email{luisgargate@ist.utl.pt}
\author{R. A. Fonseca\altaffilmark{2}}
\affil{GoLP/Instituto de Plasmas e Fus\~ao Nuclear, Instituto Superior T\'ecnico, Av. Rovisco Pais, 1049-001 Lisbon, Portugal}
\author{J. Niemiec}
\affil{Instytut Fizyki Jadrowej PAN, ul. Radzikowskiego 152, 31-342 Krak\'ow, Poland}
\author{M. Pohl\altaffilmark{3}}
\affil{Department of Physics and Astronomy, Iowa State University, Ames, IA50011, USA}
\author{R. Bingham}
\affil{SSTD, Rutherford Appleton Laboratory, Harwell Science and Innovation Campus, Didcot, Oxon, OX11 0QX UK}
\and
\author{L. O. Silva}
\affil{GoLP/Instituto de Plasmas e Fus\~ao Nuclear, Instituto Superior T\'ecnico, Av. Rovisco Pais, 1049-001 Lisbon, Portugal}
\altaffiltext{1}{Now at: Department of Astrophysical Sciences, Princeton University, NJ 08544, USA}
\altaffiltext{2}{Also at: DCTI, ISCTE - Instituto Universit\'ario de Lisboa, Av. For\c{c}as Armadas, 1649-026 Lisbon, Portugal}
\altaffiltext{3}{Now at: Institut f\"ur Physik und Astronomie, Universit\"at
Potsdam, 14476 Potsdam-Golm, Germany; and DESY, 15738 Zeuthen, Germany}

\begin{abstract}
A non-resonant instability for the amplification of the interstellar magnetic field in young Supernova Remnant (SNR) shocks was predicted by \citet{Bell:2004p24}, and is thought to be relevant for the acceleration of cosmic ray (CR) particles. For this instability, the CRs streaming ahead of SNR shock fronts drive electromagnetic waves with wavelengths much shorter than the typical CR Larmor radius, by inducing a current parallel to the background magnetic field. We explore the nonlinear regime of the non-resonant mode using Particle-in-Cell (PIC) hybrid simulations, with kinetic ions and fluid electrons, and analyze the saturation mechanism for realistic CR and background plasma parameters. In the linear regime, the observed growth rates and wavelengths match the theoretical predictions; the nonlinear stage of the instability shows a strong reaction of both the background plasma and the CR particles, with the saturation level of the magnetic field varying with the CR parameters. The simulations with CR-to-background density ratios of $n_\mathrm{CR}/n_\mathrm{b}=10^{-5}$ reveal the highest magnetic field saturation levels, with energy also being transferred to the background plasma and to the perpendicular velocity components of the CR particles. The results show that amplification factors $>10$ for the magnetic field can be achieved, and suggest that this instability is important for the generation of magnetic field turbulence, and for the acceleration of CR particles.
\end{abstract}


\section{Introduction}
Very energetic CRs ($\sim10^{14}\,\mathrm{eV}$ to $10^{15}\,\mathrm{eV}$) are thought to be accelerated in SNR shocks. There is direct evidence that electrons are accelerated up to energies of $10^{14}\,\mathrm{eV}$ at SNR sites \citep{KOYAMA:1995p201,Allen:1997p286,tanimori:1998p917,Aharonian:2001p352,Naito:1999p489,Aharonian:1999p497,Berezhko:2003p504,Vink:2003p515}, and the measured power law spectra of the CRs indicates Diffusive Shock Acceleration (DSA) as the most likely mechanism responsible for the acceleration \citep{Axford:1977p671,Bell:1978p884,Blandford:1978p891}. The acceleration of these particles up to energies of $\sim10^{15}\,\mathrm{eV}$ through the DSA mechanism requires the existence of magnetic fields much stronger than the typical $B_0\sim3\,\mathrm{\mu G}$. These strong fields have also recently been inferred from observations \citep{Longair:1994p901,Berezhko:2003p504,Vink:2003p515}; their existence, along with the requirement of stronger fields for the DSA mechanism, suggests that a magnetic field amplification mechanism is in operation.       

A possible amplification mechanism through a non-resonant instability was suggested in \citet{Bell:2004p24}, following previous work on the resonant mode in \citet{Lucek:2000p25}, and later extended to include multidimensional effects in \citet{Bell:2005p27}. The non-resonant streaming instability, part of a class of streaming instabilities derived in \citet{winskeleroy}, can be described by considering a MHD model for the background plasma, and a CR induced current imposed externally; the feedback of the electromagnetic fields on the CR particles is thus neglected. Although in the linear stage of the instability $\lambda_\mathrm{max}\ll r_\mathrm{LCR}$ ($\lambda_\mathrm{max}$ the fastest growing wavelength and $r_\mathrm{LCR}$ the typical Larmor radius of the CRs), recent full PIC simulations by \citet{Niemiec:2008p18}, \citet{Riquelme:2009p190}, \citet{reville} and \citet{stroman} indicate that the feedback mechanism of the fields on the CRs is important in the nonlinear stage, and suggest that a careful study of the instability in this regime is important to determine the saturation levels of the magnetic field. Recent works by \citet{Amato:2009p20} and \citet{Luo:2009p19}, using kinetic theory, have also shown how the saturation of this non-resonant mode depends on the details of the particle distributions.

The analysis of the behavior of the non-linear stage of the instability is very complex; full assessment of the saturation level of the magnetic field would imply a first-principle calculation of the shock formation, and the long term evolution of the shock precursor, including the self consistent acceleration of particles. The saturation of the instability was thus first assessed numerically in \citet{Bell:2004p24}, where an external current was used to model the CRs, and an MHD model was used to simulate the background plasma. The saturation mechanism found was due to the tension of the magnetic field, which grows faster than the driving term; also, the size of the simulation box was seen to limit further growth of the instability \citep{Bell:2004p24}. More recent PIC simulation results in \citet{Niemiec:2008p18} actually show a saturation level of $\delta B/B\sim1$, but consider parameters such that $\gamma_\mathrm{max}/\omega_\mathrm{ci}\ll1$ (the ratio of the growth rate of the fastest growing mode to the CR cyclotron frequency) is not strictly maintained, which is a requirement for the development of the non-resonant parallel mode. The full PIC simulation results in \citet{Riquelme:2009p190} and \citet{stroman}, taking into account the feedback of the electromagnetic fields on the CR particles, and using $m_i/m_e$ ratios up to $100$ and $n_\mathrm{CR}/n_\mathrm{b}$ down to $4\times10^{-3}$, imply a saturation level of $\delta B/B\sim10$, which occurs when the relative drift between the CRs and the background plasma decreases. Also, in \citet{reville}, similar saturation levels for the magnetic field are found. 

For PIC simulations, therefore, the instability saturates when CRs loose some of their bulk momentum to the background plasma, which starts to drift more rapidly in the direction of CRs until the velocities of the two species converge and the driver for the instability is eliminated. As shown in kinetic modeling by \citet{Luo:2009p19}, the reduction of the CR streaming motion is due to CR resonant diffusion in non-resonant magnetic turbulence, which has been recently observed in \citet{stroman}. The nonlinear evolution of the turbulence observed in  fully kinetic simulations is also in qualitative agreement with the predictions of quasi-linear calculations for non-resonant modes derived for non-relativistic beams in \citet{winskeleroy}. However, the accurate modeling of the late time evolution of the system, beyond the saturation, is limited in the kinetic simulations, when the assumption of plasma homogeneity is no longer valid. As argued in \citep{reville}, when the background plasma velocity approaches the CR bulk velocity, the density in a real scenario also changes, modifying the underlying conditions for the simulation model. This reinforces the necessity for a thorough analysis of the behavior of this instability in its non-linear stage, covering a wide range of dynamical scales, and exploring the saturation limits in different regimes.

Here, we present multi-dimensional simulation results of the non-resonant streaming instability, using the kinetic ion fluid electron hybrid model implemented in \textit{dHybrid} (see \citet{lgargate} for numerical implementation details). An important advantage of the hybrid simulations is to enable the study of the instability on the ion time scale, neglecting the high-frequency modes associated with the electrons; realistic density ratios down to $n_\mathrm{CR}/n_\mathrm{b}=10^{-5}$ can then be used, along with realistic ion-to-electron mass ratios, and simulations can be run to a point well beyond the linear stage, into the saturated state. The variation of the density ratio has an impact on the behavior of the non-resonant mode, with larger values implying the generation of different modes \citep{Bell:2004p24,Niemiec:2008p18,Riquelme:2009p190}. Though the saturation mechanism is independent of the initial linear instability \citep{stroman}, these modes lead to smaller saturation amplitudes and thus may not be directly relevant for the typical SNR shock scenarios. Also, in the hybrid simulation results shown, both the background and the CR ions are modeled kinetically, which enables the study of the feedback mechanism on the CR population. Results can then be easily compared with both the external current-driven MHD simulations, where feedback on the CR population is not modeled, and the fully kinetic simulations, which use small ion-to-electron mass ratios, and larger $n_\mathrm{CR}/n_\mathrm{b}$ ratios.

This paper is organized as follows. In section 2, the parameters used in the simulations are discussed, and the results concerning the evolution of the linear and nonlinear stages of the instability are presented. The saturation mechanism is discussed in section 3, and compared with the most recent results in the literature. Finally, in the last section, we present the conclusions and outline future research directions.

\section{Evolution of the non-resonant instability\label{section2}}
The original theoretical model of the non-resonant streaming instability, as detailed in \citet{Bell:2004p24}, considers that a magnetized background plasma with density $n_\mathrm{b}$ is stationary in the SNR shock upstream reference frame. The CR particles stream along the background magnetic field lines $B_\parallel$ with a drift velocity $v_\mathrm{sh}$, similar to the shock velocity, generating a current that drives the growth of the perpendicular components of the magnetic field $B_\perp$. For the canonical parameters in the literature \citep{Bell:2004p24}, of $B_\parallel=3\,\mathrm{\mu G}$, $n_\mathrm{b}=1\,\mathrm{cm^{-3}}$, $v_\mathrm{sh}=10000\,\mathrm{km/s}$, and $r_\mathrm{LCR}\sim1.1\times10^{13}\,\mathrm{m}$, the Alfv\'en velocity is $v_\mathrm{A}=6.6\,\mathrm{km/s}$, and the Alfv\'en Mach number is $M_\mathrm{A}=v_{sh}/v_\mathrm{A}=1515\gg 1$. Under these conditions, the reduced wave equation $\omega^2-v_\mathrm{A}^2 k^2\pm\left|B_\parallel j_\parallel \right| k/\left(n_\mathrm{b}\,m\right)=0$, where $m$ is the mass of the background ions, $j_\parallel$ is the zeroth order current imposed by the CRs, $k$ is the wavenumber and $\omega$ is the frequency of the mode, yields a purely growing mode with a growth rate $\gamma=\left[B_\parallel j_\parallel k /\left(n_\mathrm{b}\,m\right)-v_\mathrm{A}^2 k^2\right]^{1/2}$, and a maximum growing wavenumber $k_\mathrm{max}=1/2\left|B_\parallel j_\parallel\right|/(n_\mathrm{b}\,m\,v_\mathrm{A}^2)$ with a growthrate of $\gamma_\mathrm{max}=k_\mathrm{max} v_\mathrm{A}$. For the non-resonant mode to be predominant, then $1<k r_\mathrm{LCR}<\zeta v_\mathrm{sh}^2/v_\mathrm{A}^2$, with $\zeta=\left|B_\parallel j_\parallel\right| r_\mathrm{LCR}/(n_\mathrm{b}\,m\,v_\mathrm{sh}^2)$; for the canonical parameters it follows that $k_\mathrm{max}\,r_\mathrm{LCR}=3691$, and $\zeta v_\mathrm{sh}^2/v_\mathrm{A}^2=7381$. Moreover, $\gamma_\mathrm{max}/\omega_\mathrm{ci}\ll1$ is maintained, which is an essential requirement for the development of the non-resonant parallel mode.

\begin{figure} 
\figurenum{1} 
\epsscale{1.}
\plotone{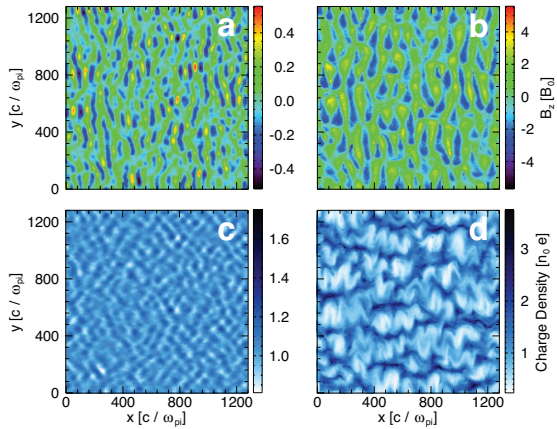}
\caption{Perpendicular magnetic field component $B_z$ (frame a  and b), and the background plasma density (frame c  and d) for run $\mathcal{B}_2$. Frames a) and c) correspond to time $7.54\,\mathrm{\gamma^{-1}}$ in the linear stage; frames b) and d) to time $16.21\,\mathrm{\gamma^{-1}}$ in the nonlinear stage.\label{fig1}} 
\end{figure}
%


The results are presented in normalized units, with the magnetic field normalized to the background field $B_0=B_\parallel$, the density normalized to the background plasma density $n_0=n_\mathrm{b}$, the velocities normalized to the Alfv\'en velocity $v_A=B_0/(n_0\,m_p\,\mu_0)^{1/2}$, the spatial dimensions normalized to the ion inertial length $c/\omega_{pi}$, and time normalized to the inverse maximum growth rate $\gamma_\mathrm{max}^{-1}$. The normalized background magnetic field in the simulations is $B_\parallel=B_x=1\,\mathrm{B_0}$, and the normalized background plasma density is $n_\mathrm{b}=1\,\mathrm{n_0}$, corresponding to SNR scenarios where $B_\parallel/n_\mathrm{b}=0.4\,\mathrm{\mu G\,cm^3}$.
\begin{deluxetable}{ccccc}
\tablecaption{Key simulation parameters\label{table1}}
\tablehead{\colhead{Run}&\colhead{Dim.}&\colhead{$n_\mathrm{CR}/n_\mathrm{b}$}&\colhead{$v_\mathrm{sh}$}&\colhead{$v_\mathrm{iso}$}}
\startdata
$\mathcal{A}$&2D&-&-&-\\
$\mathcal{B}_1$&2D&$10^{-3}$&$157\,\mathrm{v_A}$&$800\,\mathrm{v_A}$\\
$\mathcal{B}_2$&2D&$10^{-4}$&$1570\,\mathrm{v_A}$&$800\,\mathrm{v_A}$\\
$\mathcal{B}_3$&2D&$10^{-5}$&$15700\,\mathrm{v_A}$&$800\,\mathrm{v_A}$\\
$\mathcal{C}_1$&2D&$10^{-3}$&$157\,\mathrm{v_A}$&$160\,\mathrm{v_A}$\\
$\mathcal{C}_2$&2D&$10^{-3}$&$157\,\mathrm{v_A}$&$320\,\mathrm{v_A}$\\
$\mathcal{C}_3$&2D&$10^{-3}$&$157\,\mathrm{v_A}$&$500\,\mathrm{v_A}$\\
$\mathcal{D}$&3D&$10^{-3}$&$157\,\mathrm{v_A}$&$800\,\mathrm{v_A}$\\
\enddata
\end{deluxetable}
The main parameters for the simulations are described in Table \ref{table1}. The CRs are modeled through a constant external current in run $\mathcal{A}$, and with kinetic particles in all other runs (the kinetic-CR runs). The value of the initial CR current is always $j=n_\mathrm{CR}\,e\,v_\mathrm{sh}=\pi/20\,\left[\mathrm{n_0\,e\,v_A}\right]$ (where $\left[\mathrm{n_0\,e\,v_A}\right]$ are the normalized current density units), which yields a maximum growing wavelength $\lambda_\mathrm{max}=80\,\mathrm{c/\omega_{pi}}$ ($k_\mathrm{max}=\pi/40\,\mathrm{\omega_{pi}/c}$). For the kinetic-CR runs, the total velocity in the CR population is initially set to $\vec{v}_\mathrm{sh}+\vec{v}_\mathrm{iso}$, with $\vec{v}_\mathrm{iso}$ the isotropic velocity component, so that the whole CR population drifts with the shock velocity, as observed in the upstream reference frame. For runs $\mathcal{B}_1$ through $\mathcal{B}_3$ the CR-to-backghround density ratio was varied, along with the drift velocity $v_\mathrm{sh}$, and for runs $\mathcal{C}_1$ through $\mathcal{C}_3$, the $v_\mathrm{iso}$ was varied, enabling the analysis of the dependence of the instability on the scale separation between the typical Larmor radius of the CRs, and $\lambda_\mathrm{max}$. Finally, in run $\mathcal{D}$, the parameters from run $\mathcal{B}_1$ were used in a 3D setup. For all the runs the time step was chosen to be $\Delta t=0.001\,\mathrm{\omega_{ci}^{-1}}$, as to account for the cavitation in the background density plasma, since low density zones can render the hybrid (and MHD) algorithms unstable \citep{lipatov,lgargate}. Full convergence tests have been performed to assure that the saturation level of the magnetic field was not affected by the choice of the time step.

The simulation box size is $L_x=L_y=1280\,\mathrm{c/\omega_{pi}}$ for the 2D runs, and $L_z=640\,\mathrm{c/\omega_{pi}}$ for the third dimension in the 3D run presented. The $L_x$ and $L_y$ dimensions are then $\sim16\times\lambda_\mathrm{max}$, corresponding to $L_x/L_\mathrm{diff}\sim1$, where $L_\mathrm{diff}\sim D/v_\mathrm{sh}$ is the CR diffusion length scale in the Bohm diffusion limit for the typical CR velocities used in the simulations. This choice of parameters then allows for the CR Larmor radius to be fully resolved, and is such that the non-resonant mode wavelength can grow to be of the order of  $r_\mathrm{LCR}$ in the non-linear regime, enabling the analysis of the back-reaction of the generated electromagnetic fields on the CR particles.

The diverse parameters chosen for the four sets of runs ($\mathcal{A}$ through $\mathcal{D}$) does not directly reflect typical physical conditions found in young SNR shock precursors for all the runs; instead a wide dynamical range of parameters was analyzed. The results of run $\mathcal{A}$ are most directly comparable to the simulations by \citet{Bell:2004p24}, in which the CR current was imposed externally, thus excluding any feedback of magnetic turbulence on theCRs.
For the $\mathcal{B}$ set of runs, the driving current is maintained constant while the driving energy of the CR beam is increased from run $\mathcal{B}_1$ to run $\mathcal{B}_3$ by increasing $v_\mathrm{sh}$ and decreasing $n_\mathrm{CR}$; as such $v_\mathrm{iso}/v_\mathrm{sh}$ is smaller than the typical values found in young SNR shock precursors, especially for runs $\mathcal{B}_2$ and $\mathcal{B}_3$. The $\mathcal{C}$ set of runs explores the behavior of the instability when progressing from the non-resonant to the resonant limit of the CR streaming instability by lowering the $v_\mathrm{iso}$ velocity and thus the $r_\mathrm{LCR}/\lambda_\mathrm{max}$ ratio. 

Comparison of results between 3D and 2D runs showed no significant difference in the development of the instability. This motivated the analysis of the system using a 2D simulation setup. For run $\mathcal{D}$, the measured wavelength of the instability was $\lambda=80\,\mathrm{c/\omega_{pi}}$, and the growth rate $\gamma\sim0.889$ of the non-resonant mode is the same as for all the runs in the $\mathcal{B}$ set. Figure \ref{fig1} shows the evolution of the $B_z$ magnetic field component and the density of the background plasma for run $\mathcal{B}_2$; Fig. \ref{fig1} a) and c) correspond to the linear stage and show the well defined wavelength $\lambda=80\,\mathrm{c/\omega_{pi}}$ in the magnetic field, and a low amplitude modulation of the background density. In the nonlinear stage, cavities are formed in the background plasma, due to turbulent plasma motions induced by the electromagnetic waves being generated. The well defined wave structure of the linear stage disappears in Fig. \ref{fig1} b), with magnetic field compression zones being formed, which are correlated with the density enhancements at the same positions, Fig. \ref{fig1} d), confirming the results of \citep{Riquelme:2009p190} for lower mass ratios. Also, very similar results are observed for runs $\mathcal{B}_1$ and $\mathcal{B}_3$.
\begin{figure} 
\figurenum{2} 
\epsscale{1.}
\plotone{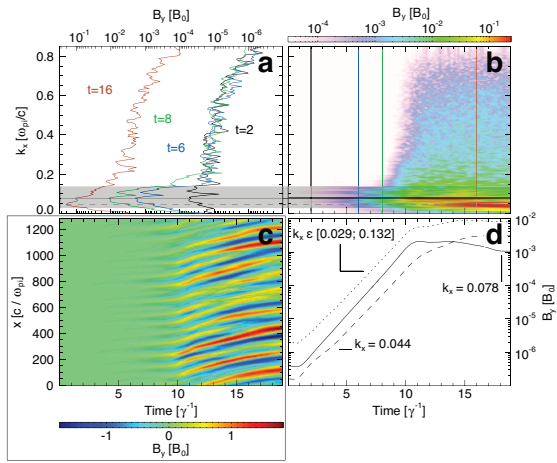}
\caption{Evolution of the perpendicular magnetic field component $B_y$ in time for run $\mathcal{B}_2$. Frame a) shows lineouts of the power distribution in k space (frame b) at times: $2\,\mathrm{\gamma^{-1}}$ (black line), $6\,\mathrm{\gamma^{-1}}$ (blue line), $8\,\mathrm{\gamma^{-1}}$ (green line), and $16\,\mathrm{\gamma^{-1}}$ (red line). Frame d) shows the growth rates for $k_x=0.078\,\mathrm{\omega_{pi}/c}$ (solid line), $k_x=0.044\,\mathrm{\omega_{pi}/c}$ (dashed line), and for the $0.029<k_x<0.132\,\mathrm{\omega_{pi}/c}$ high-growth band (dotted line). Frame b) is obtained by fourier transforming frame c) in the spatial dimension.\label{fig2}} 
\end{figure}

In Fig. \ref{fig2}, the main characteristics of the instability that agree with expectations of a quasi-linear theory are observable. At $t\sim9\,\mathrm{\gamma^{-1}}$ the instability becomes nonlinear, and saturation of the linear mode is observed a short time after, around $t=10\,\mathrm{\gamma^{-1}}$ (Fig. \ref{fig2} b) and d). In the linear stage, the growth rate from Fig. \ref{fig2} d) yields $\gamma=.889$ which deviates from the theoretical value by $11\%$. It is also clear from Fig. \ref{fig2} a) that the peak power is for $k\sim0.078\,\mathrm{\omega_{pi}/c}$ up to $t=8\,\mathrm{\gamma^{-1}}$ (dotted line), and that at $t=16\,\mathrm{\gamma^{-1}}$ the peak shifts to $k\sim0.044\,\mathrm{\omega_{pi}/c}$ (dashed line).

In the nonlinear stage, Fig. \ref{fig2} b) and c), the increase in power over large $k$ is consistent with the magnetic field structures in Fig. \ref{fig1} b), and indicates an increase in the turbulence level in $B_z$. Figure \ref{fig2} c) also shows that the waves move in the $x$ direction at a velocity $v\sim1.8\,\mathrm{v_A}$ in the nonlinear stage.

\section{Saturation mechanism\label{section3}}
\begin{figure} 
\figurenum{3} 
\epsscale{1.}
\plotone{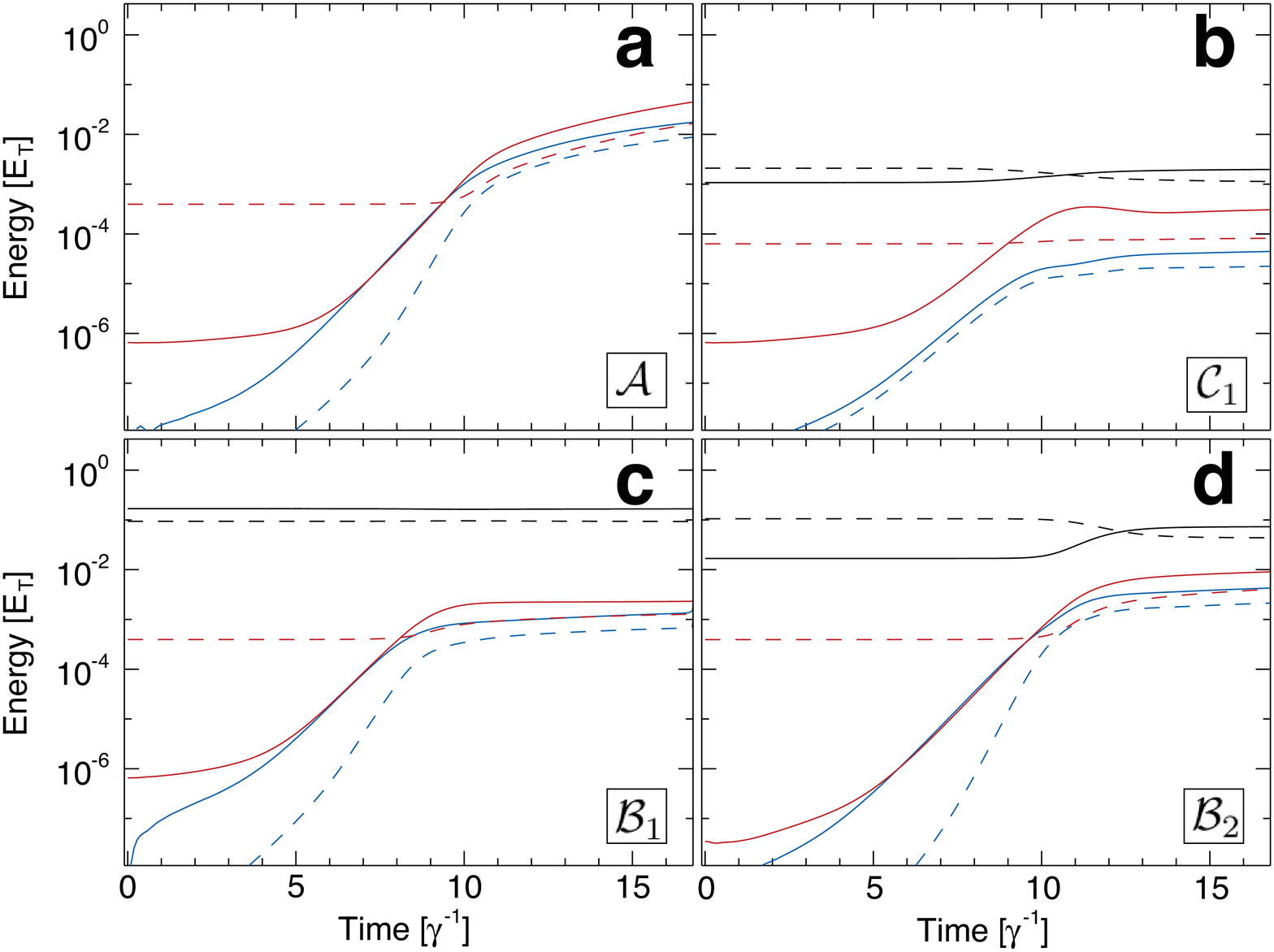}
\caption{Time evolution of the energy for the magnetic field (red lines), the CRs (black lines), and the background plasma (blue lines); dashed lines represent the component parallel to the background magnetic field $x$, and the solid lines represent the perpendicular component. Frame a) shows the evolution for run $\mathcal{A}$, frame b) shows the evolution for run $\mathcal{C}_1$, frame c) shows the evolution for run $\mathcal{B}_1$, and frame d) shows the evolution for run $\mathcal{B}_2$. The energy scale is normalized to the total energy in run $\mathcal{B}_3$.\label{fig3}} 
\end{figure}

The maximum energy level in the perpendicular magnetic field components, at the end of the linear stage, is similar for the runs $\mathcal{B}_1$, $\mathcal{B}_2$ (see Fig. \ref{fig3}), and $\mathcal{B}_3$. In the linear stage of the instability, the behavior of run $\mathcal{A}$ is identical to the kinetically-driven runs, $\mathcal{B}_1$ and $\mathcal{B}_2$. In the nonlinear stage, however, the magnetic field energy peaks at $t=10\,\mathrm{\gamma^{-1}}$ for runs $\mathcal{B}_1$ and $\mathcal{B}_2$, while for run $\mathcal{A}$ the magnetic field energy still grows for $t>10\,\mathrm{\gamma^{-1}}$. The growth beyond $t=10\,\mathrm{\gamma^{-1}}$ for run $\mathcal{A}$ is due to the continuous injection of energy into the system; a similar growth is observed for run $\mathcal{B}_3$, at a slower rate, which is due to the excess of free energy in the CR population for this run. Finally, run $\mathcal{C}_1$ (Fig. \ref{fig3} b) represents an hybrid scenario, between the resonant and the non-resonant modes of the instability ($r_\mathrm{LRC}/\lambda_\mathrm{max}\sim1.63$), and thus the magnetic field growth rate is smaller in the linear stage.

The CR population velocity distribution is nearly isotropic at the end of all the simulations (as seen in the upstream / simulation frame). In the linear stage, the growth rates are identical between runs in the non-resonant regime, and the background plasma also gains energy at the same rate in the perpendicular direction (Fig. \ref{fig3} c) and d). Unlike what is predicted from theory, $v_x$ of the background plasma increases, and the parallel energy becomes comparable to the perpendicular energy. With this energy increase, the plasma is no longer magnetized, and thus the nature of the non-resonant mode changes. 
\begin{figure} 
\figurenum{4} 
\epsscale{1.}
\plotone{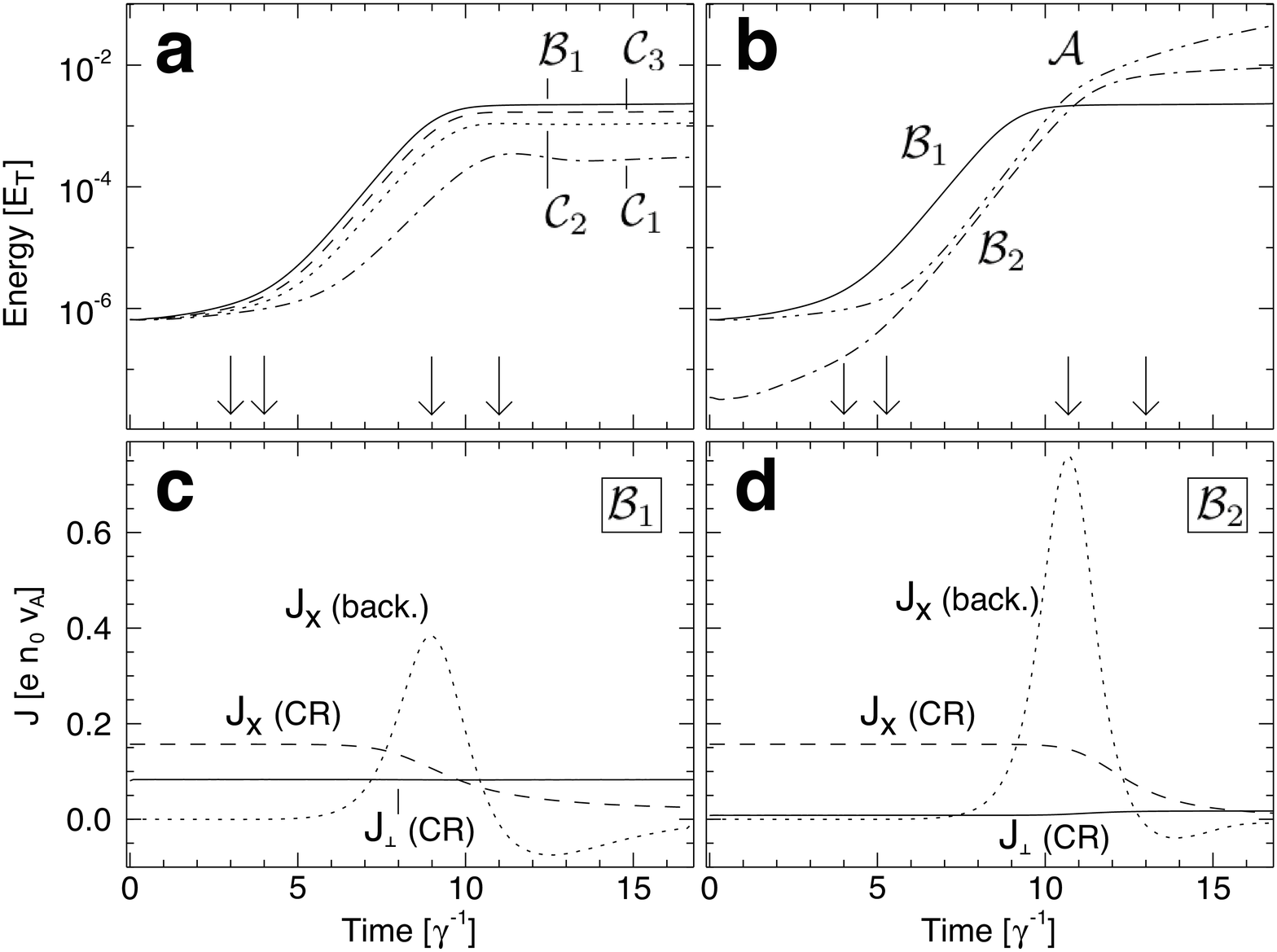}
\caption{Time evolution of the energy in the perpendicular components of the magnetic field (top frames), and the time evolution of the current density for the ion component of the CRs, and the ion component of the background plasma (lower frames). Frame c) shows the current density components for run $\mathcal{B}_1$, and frame d) shows the current density components for run $\mathcal{B}_2$; the vertical arrows, from left to right, indicate points where the background plasma current, $J_x$ is: at a minimum, zero, at a maximum, and zero, respectively. Lines are labeled in the plot for ease of interpretation. The energy scale is normalized to the total energy in run $\mathcal{B}_3$.\label{fig4}}
\end{figure}

In order to understand the dependence of the non-resonant mode on the scale-separation between $\lambda_\mathrm{max}$ and $r_\mathrm{LCR}$, the isotropic velocity was varied in runs $\mathcal{C}_1$ through $\mathcal{C}_3$. The typical CR Larmor radius varies as $r_\mathrm{LCR}=\left\{130.6,261.3,408.2\right\}\,\mathrm{c/\omega_{pi}}$ (and $653.2\,\mathrm{c/\omega_{pi}}$ for run $\mathcal{B}_1$). Figure \ref{fig3} and Fig. \ref{fig4} a) and b) show that for $r_\mathrm{LCR}/\lambda_\mathrm{max}$ ratios greater than $\sim3.2$ (run $\mathcal{C}_2$), the behavior is consistent with the non-resonant mode, and only run $\mathcal{C}_1$ ($r_\mathrm{LCR}/\lambda_\mathrm{max}=1.63$) results in an hybrid scenario. Analysis of Fig. \ref{fig4} c) and d) also shows that for runs $\mathcal{B}_1$ and $\mathcal{B}_2$, at $t\sim9\,\mathrm{\gamma^{-1}}$, the $J_x$ current density component for the CRs is not close to zero. This means that the saturation mechanism does not depend directly on the amount of free energy in the CR population. Instead, the saturation of the instability is due to the energy gain of the background plasma, which results in an average bulk flow in the CR propagation direction, and in the de-magnetization of the plasma. 

The relation between $J_x$ for the background plasma, and both the onset and the peak in the magnetic field energy level of the instability is highlighted in Fig. \ref{fig4} c) and d). The peak of $J_x$, for the background plasma, marks the end of the linear stage of the instability. Likewise, when $J_x$ is at a minimum ($t\sim3\,\mathrm{\gamma^{-1}}$), the instability is entering the initial linear stage. Finally, for run $\mathcal{B}_1$, Fig. \ref{fig4} a) and c), the energy in the CRs at the saturation point is transferred to the background plasma and to the magnetic field; for run $\mathcal{B}_2$, Fig. \ref{fig4} b) and d), some of the energy is converted into perpendicular energy in the CR population, as $J_\perp$ increases from $t\sim10.8\,\mathrm{\gamma^{-1}}$ onwards. 

\section{Discussion and conclusions}
The identification of the nonlinear saturation mechanism for the non-resonant streaming instability is important for the determination of the maximum magnetic field amplification in SNR shock scenarios. Amplification factors of $B_\perp/B_\parallel\sim10$, with local temporary peaks of $B_\perp/B_\parallel\sim25$ are observable in the current hybrid simulations; similar results were also recently shown in the kinetic simulations of \citet{Riquelme:2009p190} \citet{reville} and \citet{stroman}, with reduced temporal and spatial scales, mass ratios and density ratios. The hybrid simulation results presented expand the current knowledge of the non-resonant streaming instability by extending the dynamical range under study, and by performing a detailed study of the instability under different driving regimes.

Leveraging on the hybrid simulations presented here, it is possible to scan the parameter space and analyze the nonlinear stage of the instability in detail, using realistic $n_\mathrm{CR}/n_\mathrm{b}$ and $m/m_e$ ratios. Close examination of Fig. \ref{fig3} and Fig. \ref{fig4} shows a number of important points, relating to the saturation mechanism. As the $n_\mathrm{CR}/n_\mathrm{b}$ ratio is decreased towards $10^{-5}$, the energy in the CRs increases, as in our setup the current is initialized at the same level in all simulations; the increase in the free energy of the CRs, associated with the higher drift velocity $v_\mathrm{sh}$, does not change the peak energy level of the magnetic field. The excess free energy is instead transferred to the background plasma, and also to the perpendicular velocity components of the CR population (Fig. \ref{fig3} b), run $\mathcal{C}_1$).

The reaction of the background plasma is critical in the nonlinear stage of the instability. The average local Alfv\'en velocity ($v_A\propto B_\parallel/\sqrt{n}$ calculated in each cell and averaged over the entire simulation box) is approximately constant over the linear development of the instability. In the nonlinear stage, $v_A$ increases with the local magnetic field $B_\parallel$, and with the background density variations (which are in phase), and thus $v_A\propto\sqrt{\delta B/B_\parallel}$. The background plasma accelerates in the CR propagation direction, and the current density peaks, marking the end of the linear stage (Fig. \ref{fig4}). At this point in time, the CRs $v_x$ velocity component is still $v_x\sim1500\,\mathrm{v_A}$, well above the Alfv\'en velocity and above the background plasma drift velocity, which is sub-Alfv\'enic. This shows that the saturation is not dependent on the amount of free energy in the CRs, but instead results from the reduction of the scale separation of the background plasma, and the CRs.  

The CR's $J_\perp$ increases for run $\mathcal{B}_2$, Fig. \ref{fig4} d), as it is substantially lower than $J_x$ initially, and the distribution becomes isotropic after the saturation, at $t\sim15\,\mathrm{\gamma^{-1}}$. For run $\mathcal{B}_1$ in Fig. \ref{fig4} c), the distribution is also isotropic at the end, but the CRs $J_\perp$ velocity component is not affected. Increasing the free energy in the CRs does not affect the saturation level of the magnetic field significantly, as long as the driving current is maintained. 

The hybrid simulation results presented thus show that, for a broad range of parameters, the nonlinear growth of the streaming instability is independent on the energy of the driving CR population, and depends only on the current carried by the CRs. The amplification factor of $\sim10$ for the perpendicular magnetic field above the seed $B_\parallel$ feed indicates that the mechanism is relevant for the magnetic field amplification by CR particles in SNR shocks. Beyond the saturation point, for $t>16\,\mathrm{\gamma^{-1}}$, the CR particle distribution becomes isotropic, with the instability saturating after 10 e-foldings. The de-magnetization of the background plasma, as it gains energy, hinders further growth of the non-resonant mode; when both the CRs and the background plasma are unmagnetized, the instability should behave more like the Weibel instability \citep{weibel}.
 
Our results thus indicate that the instability begins to saturate when the current carried by the background ions is similar to the current carried by the CR population (see Fig. \ref{fig4}). This results in a final average velocity for the background plasma of $v_\mathrm{b}\sim n_\mathrm{CR}/n_\mathrm{b} v_\mathrm{sh}$ so that $V_\mathrm{b}\sim10^{-5}v_\mathrm{sh}$ for realistic parameters. At this point, the instability is in the non-linear stage, and the magnetic field growth rate is much lower ($t>9\,\mathrm{\gamma^{-1}}$ in Fig. \ref{fig3} c) and Fig \ref{fig4} a) and c) for run $\mathcal{B}_1$). In fact, even in the linear stage of the instability, energy is being transferred to the perpendicular components of the background plasma at a greater rate than into the magnetic field perpendicular components (see Fig. \ref{fig3} c), and thus at some point the background plasma de-magnetizes and the instability saturates.

Previous simulation PIC results \citep{Riquelme:2009p190,reville,stroman} show similar saturation levels for the magnetic field $\delta B/B\sim10$: the instability saturates when the background plasma is moving with a bulk velocity comparable to the CR population drift speed, which is equivalent to saying that the currents carried by the background plasma and the CRs are similar (since $n_\mathrm{CR}\sim n_\mathrm{b}$, and thus $J_\mathrm{CR}=n_\mathrm{CR}\,e\,v_\mathrm{sh}\sim J_\mathrm{b}=n_\mathrm{b}\,e\,v_\mathrm{b}$). It has been claimed \citep{reville} that this saturation limit might be numerical rather than physical because in a real shock precursor a significant variation in the background plasma velocity should be concurrent with a variation in the background plasma density.However, here we show for $m_i\gg m_e$ and $n_\mathrm{CR}\ll n_\mathrm{b}$ that the instability saturates due to the de-magnetization of the background plasma; at that point the currents carried by the background ions and the CRs are comparable, although the velocities differ significantly (since in our hybrid simulations $n_\mathrm{CR}\ll n_\mathrm{b}$ implies that $v_\mathrm{b}\sim v_\mathrm{sh}\,n_\mathrm{CR}/n_\mathrm{b}$ from the equality of the currents at the saturation). Our simulations thus reinforce the result $\delta B/B\sim10$ at saturation, and indicate that the instability should be important in a young SNR shock precursor. The simulations present in the literature make assumptions about the physical scenarios, and therefore significant care should be taken when interpreting and comparing the results. For instance, the back-reaction of the fields in the CRs is not accounted for when using external currents as drivers, in full PIC simulations mass ratios and density ratios far from the physical conditions have been explored, and the dynamics of electrons is not accounted for in hybrid simulations. Thus, our results further indicate that first-principle simulations of SNR shocks and particle acceleration should be attempted in a future work, since the different simulation methods can capture the instability, but differences might still be found when the simulation setup and the models are refined in order to more closely match the experimental conditions.

Further analysis for SNR shocks will be possible by considering the saturation mechanism when unperturbed CRs are continuously injected into the simulation box. This will allow for an improved comparison of the SNR shock scenario, where CR particles are also thought to be continuously injected from the shock into the upstream medium. This will be explored in a future publication, leveraging on the unique characteristics of the hybrid model. Finally, with hybrid simulations, it will also be possible to determine the dependence of usual characteristics of the instability with the distance to the shock, and to study in detail the energy profile of the accelerated CR particles.

\section{Acknowledgments}
This work was supported by Funda\c{c}\~ao para a Ci\^encia e a Tecnologia (FCT/Portugal) under grant SFRH/BD/17750/2004, and grant POCI/FIS/66823/2006. The simulations presented in this paper were produced using the IST Cluster (IST/Portugal). The work of JN is supported by MNiSW research project N203 393034, and the Foundation for Polish Science through the HOMING program, which is supported by a grant from Iceland, Liechtenstein, and Norway through the EEA Financial Mechanism. The authors also want to thank KITP (UCSB) where part of this work was done, partially supported by NSF under Grant No. PHY05-51164, and A. R. Bell for the useful discussions on the instability mechanism.

\bibliographystyle{apj}
\bibliography{bibliography}
\end{document}